\pdfoutput=1

\documentclass[12pt,a4paper]{article}

\usepackage{ifthen} 
\newboolean{pdflatex}
\setboolean{pdflatex}{true} 

\newboolean{articletitles}
\setboolean{articletitles}{true} 

\newboolean{uprightparticles}
\setboolean{uprightparticles}{false} 

\newboolean{inbibliography}
\setboolean{inbibliography}{false} 

\RequirePackage[normalem]{ulem} 
\RequirePackage{color}\definecolor{RED}{rgb}{1,0,0}\definecolor{BLUE}{rgb}{0,0,1} 

\def\paperauthors{LHCb collaboration} 
\def\paperasciititle{First observation of the doubly charmed baryon decay Xi_cc++ -> Xi_c+ pi+} 
\def\papertitle{First observation of the doubly charmed baryon decay \XiccppXicDecay} 
\def\paperkeywords{{High Energy Physics}, {LHCb}} 
\def\papercopyright{\the\year\ CERN for the benefit of the LHCb collaboration} 
\def\paperlicence{CC-BY-4.0 licence}
\def\paperlicenceurl{https://creativecommons.org/licenses/by/4.0/}

\usepackage[top=1in, bottom=1.25in, left=1in, right=1in]{geometry}

\usepackage{microtype}
\usepackage{lineno}  
\usepackage{xspace} 
\usepackage{caption} 

\usepackage{graphicx}  
\usepackage{color}
\usepackage{colortbl}
\graphicspath{{./figs/}} 

\usepackage{amsmath} 
\usepackage{amssymb}
\usepackage{amsfonts}
\usepackage{upgreek} 

\newcommand*\patchAmsMathEnvironmentForLineno[1]{%
\expandafter\let\csname old#1\expandafter\endcsname\csname #1\endcsname
\expandafter\let\csname oldend#1\expandafter\endcsname\csname
end#1\endcsname
 \renewenvironment{#1}%
   {\linenomath\csname old#1\endcsname}%
   {\csname oldend#1\endcsname\endlinenomath}%
}
\newcommand*\patchBothAmsMathEnvironmentsForLineno[1]{%
  \patchAmsMathEnvironmentForLineno{#1}%
  \patchAmsMathEnvironmentForLineno{#1*}%
}
\AtBeginDocument{%
\patchBothAmsMathEnvironmentsForLineno{equation}%
\patchBothAmsMathEnvironmentsForLineno{align}%
\patchBothAmsMathEnvironmentsForLineno{flalign}%
\patchBothAmsMathEnvironmentsForLineno{alignat}%
\patchBothAmsMathEnvironmentsForLineno{gather}%
\patchBothAmsMathEnvironmentsForLineno{multline}%
\patchBothAmsMathEnvironmentsForLineno{eqnarray}%
}

\usepackage{hyperxmp}

\usepackage[pdftex,
            pdfauthor={\paperauthors},
            pdftitle={\paperasciititle},
            pdfkeywords={\paperkeywords},
            pdfcopyright={Copyright (C) \papercopyright},
            pdflicenseurl={\paperlicenceurl}]{hyperref}

\usepackage[all]{hypcap} 

\usepackage{xspace} 
\usepackage{upgreek}

\def\lhcb {\mbox{LHCb}\xspace}

\def\MagUp {\mbox{\em Mag\kern -0.05em Up}\xspace}

\ifthenelse{\boolean{uprightparticles}}%
{

 \def\Pmu         {\ensuremath{\upmu}\xspace}

 \def\Ppi         {\ensuremath{\uppi}\xspace}

 \def\Ppsi        {\ensuremath{\uppsi}\xspace}

 \def\PDelta      {\ensuremath{\Delta}\xspace}                 
 \def\PXi      {\ensuremath{\Xi}\xspace}                 
 \def\PLambda      {\ensuremath{\Lambda}\xspace}                 
 \def\PSigma      {\ensuremath{\Sigma}\xspace}                 
 \def\POmega      {\ensuremath{\Omega}\xspace}                 
 \def\PUpsilon      {\ensuremath{\Upsilon}\xspace}                 
 
 \def\PB      {\ensuremath{\mathrm{B}}\xspace}                 
                  
 \def\PD      {\ensuremath{\mathrm{D}}\xspace}

 \def\PJ      {\ensuremath{\mathrm{J}}\xspace}                 
 \def\PK      {\ensuremath{\mathrm{K}}\xspace}

 \def\Pb      {\ensuremath{\mathrm{b}}\xspace}                 
 \def\Pc      {\ensuremath{\mathrm{c}}\xspace}

 \def\Pi      {\ensuremath{\mathrm{i}}\xspace}

 \def\Pp      {\ensuremath{\mathrm{p}}\xspace}

}
{

 \def\Pmu         {\ensuremath{\mu}\xspace}

 \def\Ppi         {\ensuremath{\pi}\xspace}

 \def\Ppsi        {\ensuremath{\psi}\xspace}                 
                  
 \mathchardef\PDelta="7101
 \mathchardef\PXi="7104
 \mathchardef\PLambda="7103
 \mathchardef\PSigma="7106
 \mathchardef\POmega="710A
 \mathchardef\PUpsilon="7107
                  
 \def\PB      {\ensuremath{B}\xspace}                 
                  
 \def\PD      {\ensuremath{D}\xspace}

 \def\PJ      {\ensuremath{J}\xspace}                 
 \def\PK      {\ensuremath{K}\xspace}

 \def\Pb      {\ensuremath{b}\xspace}                 
 \def\Pc      {\ensuremath{c}\xspace}

 \def\Pi      {\ensuremath{i}\xspace}

 \def\Pp      {\ensuremath{p}\xspace}

}

\makeatletter
\ifcase \@ptsize \relax
  \newcommand{\miniscule}{\@setfontsize\miniscule{4}{5}}
\or
  \newcommand{\miniscule}{\@setfontsize\miniscule{5}{6}}
\or
  \newcommand{\miniscule}{\@setfontsize\miniscule{5}{6}}
\fi
\makeatother

\DeclareRobustCommand{\optbar}[1]{\shortstack{{\miniscule (\rule[.5ex]{1.25em}{.18mm})}
  \\ [-.7ex] $#1$}}

\def\mumu       {{\ensuremath{\Pmu^+\Pmu^-}}\xspace}

\def\cquark    {{\ensuremath{\Pc}}\xspace}

\def\bquark    {{\ensuremath{\Pb}}\xspace}

\def\pion   {{\ensuremath{\Ppi}}\xspace}

\def\pip    {{\ensuremath{\pion^+}}\xspace}
\def\pim    {{\ensuremath{\pion^-}}\xspace}

\def\kaon    {{\ensuremath{\PK}}\xspace}
  \def\Kbar    {{\kern 0.2em\overline{\kern -0.2em \PK}{}}\xspace}

\def\KorKbar    {\kern 0.18em\optbar{\kern -0.18em K}{}\xspace}

\def\Kp      {{\ensuremath{\kaon^+}}\xspace}
\def\Km      {{\ensuremath{\kaon^-}}\xspace}

\def\Dbar    {{\kern 0.2em\overline{\kern -0.2em \PD}{}}\xspace}

\def\DorDbar    {\kern 0.18em\optbar{\kern -0.18em D}{}\xspace}

\def\B       {{\ensuremath{\PB}}\xspace}
\def\Bbar    {{\ensuremath{\kern 0.18em\overline{\kern -0.18em \PB}{}}}\xspace}

\def\BorBbar    {\kern 0.18em\optbar{\kern -0.18em B}{}\xspace}

\def\Bu      {{\ensuremath{\B^+}}\xspace}

\def\Bp      {{\ensuremath{\Bu}}\xspace}

\def\jpsi     {{\ensuremath{{\PJ\mskip -3mu/\mskip -2mu\Ppsi\mskip 2mu}}}\xspace}

  \def\Y#1S{\ensuremath{\PUpsilon{(#1S)}}\xspace}

\def\proton      {{\ensuremath{\Pp}}\xspace}

\def\Xires       {{\ensuremath{\PXi}}\xspace}

\def\Lz          {{\ensuremath{\PLambda}}\xspace}
\def\Lbar        {{\ensuremath{\kern 0.1em\overline{\kern -0.1em\PLambda}}}\xspace}
\def\LorLbar    {\kern 0.18em\optbar{\kern -0.18em \PLambda}{}\xspace}

\def\Lb      {{\ensuremath{\Lz^0_\bquark}}\xspace}

\def\Lc      {{\ensuremath{\Lz^+_\cquark}}\xspace}

\def\Xicp    {{\ensuremath{\Xires^+_\cquark}}\xspace}

\def\Xiccpp    {{\ensuremath{\Xires^{++}_{\cquark\cquark}}}\xspace}

\def\BF         {{\ensuremath{\mathcal{B}}}\xspace}

\def\to                 {\ensuremath{\rightarrow}\xspace}

\def\eps   {{\ensuremath{\varepsilon}}\xspace}

\def\AT#1     {\ensuremath{A_{\mathrm{T}}^{#1}}\xspace}           

\def\C#1      {\ensuremath{\mathcal{C}_{#1}}\xspace}                       
\def\Cp#1     {\ensuremath{\mathcal{C}_{#1}^{'}}\xspace}                    
\def\Ceff#1   {\ensuremath{\mathcal{C}_{#1}^{\mathrm{(eff)}}}\xspace}        
\def\Cpeff#1  {\ensuremath{\mathcal{C}_{#1}^{'\mathrm{(eff)}}}\xspace}       
\def\Ope#1    {\ensuremath{\mathcal{O}_{#1}}\xspace}                       
\def\Opep#1   {\ensuremath{\mathcal{O}_{#1}^{'}}\xspace}                    

\newcommand{\tev}{\ifthenelse{\boolean{inbibliography}}{\ensuremath{~T\kern -0.05em eV}}{\ensuremath{\mathrm{\,Te\kern -0.1em V}}}\xspace}
\newcommand{\gev}{\ensuremath{\mathrm{\,Ge\kern -0.1em V}}\xspace}
\newcommand{\mev}{\ensuremath{\mathrm{\,Me\kern -0.1em V}}\xspace}
\newcommand{\kev}{\ensuremath{\mathrm{\,ke\kern -0.1em V}}\xspace}
\newcommand{\ev}{\ensuremath{\mathrm{\,e\kern -0.1em V}}\xspace}
\newcommand{\gevc}{\ensuremath{{\mathrm{\,Ge\kern -0.1em V\!/}c}}\xspace}
\newcommand{\mevc}{\ensuremath{{\mathrm{\,Me\kern -0.1em V\!/}c}}\xspace}
\newcommand{\gevcc}{\ensuremath{{\mathrm{\,Ge\kern -0.1em V\!/}c^2}}\xspace}
\newcommand{\gevgevcccc}{\ensuremath{{\mathrm{\,Ge\kern -0.1em V^2\!/}c^4}}\xspace}
\newcommand{\mevcc}{\ensuremath{{\mathrm{\,Me\kern -0.1em V\!/}c^2}}\xspace}

\def\invfb   {\ensuremath{\mbox{\,fb}^{-1}}\xspace}

\def\ps   {\ensuremath{{\mathrm{ \,ps}}}\xspace}

\newcommand{\stat}{\ensuremath{\mathrm{\,(stat)}}\xspace}
\newcommand{\syst}{\ensuremath{\mathrm{\,(syst)}}\xspace}

\newcommand{\chisq}{\ensuremath{\chi^2}\xspace}

\newcommand{\chisqip}{\ensuremath{\chi^2_{\text{IP}}}\xspace}

\def\gsim{{~\raise.15em\hbox{$>$}\kern-.85em
          \lower.35em\hbox{$\sim$}~}\xspace}
\def\lsim{{~\raise.15em\hbox{$<$}\kern-.85em
          \lower.35em\hbox{$\sim$}~}\xspace}

\def\sqs   {\ensuremath{\protect\sqrt{s}}\xspace}

\def\pt         {\mbox{$p_{\mathrm{ T}}$}\xspace}

\def\mrad{\ensuremath{\mathrm{ \,mrad}}\xspace}

\def\evtgen     {\mbox{\textsc{EvtGen}}\xspace}

\def\geant      {\mbox{\textsc{Geant4}}\xspace}

\def\photos     {\mbox{\textsc{Photos}}\xspace}

\def\pythia     {\mbox{\textsc{Pythia}}\xspace}

\def\tell1  {TELL1\xspace}
\def\ukl1   {UKL1\xspace}

\newcommand{\eg}{\mbox{\itshape e.g.}\xspace}

\usepackage{cite} 
\usepackage{mciteplus}

\usepackage{longtable} 

\usepackage{xspace} 
\usepackage{upgreek}

\def\DeltaOrDeltabar  {\kern 0.18em\optbar{\kern -0.18em \PDelta}{}\xspace}
\def\XiOrXibar        {\kern 0.18em\optbar{\kern -0.18em \PXi}{}\xspace}
\def\SigmaOrSigmabar  {\kern 0.18em\optbar{\kern -0.18em \PSigma}{}\xspace}
\def\OmegaOrOmegabar  {\kern 0.18em\optbar{\kern -0.18em \POmega}{}\xspace}

\def\Lcp          {{\ensuremath{\Lz^+_\cquark}}\xspace}

\def\Xicp         {{\ensuremath{\Xires^+_\cquark}}\xspace}

\def\Xiccpp       {{\ensuremath{\Xires^{++}_{\cquark\cquark}}}\xspace}

\def\XiccppFullDecay{{\ensuremath{\Xiccpp\to\Lc(\to \proton \Km \pip) \, \Km\pip\pip}}\xspace}
\def\XiccppXicFullDecay{{\ensuremath{\Xiccpp\to\Xicp(\to \proton \Km \pip) \, \pip}}\xspace}
\def\LbLconepi{{\ensuremath{\Lb\to\Lcp(\to \proton \Km \pip)\pim}}\xspace}
\def\LbLcthreepi{{\ensuremath{\Lb\to\Lcp(\to \proton \Km \pip)\pim\pip\pim}}\xspace}

\newcommand{\TeVnosp}{\ifthenelse{\boolean{inbibliography}}{\ensuremath{~T\kern -0.05em eV}}{\ensuremath{\mathrm{Te\kern -0.1em V}}}}
\newcommand{\GeVnosp}{\ensuremath{\mathrm{Ge\kern -0.1em V}}}
\newcommand{\MeVnosp}{\ensuremath{\mathrm{Me\kern -0.1em V}}}
\newcommand{\keVnosp}{\ensuremath{\mathrm{ke\kern -0.1em V}}}
\newcommand{\eVnosp}{\ensuremath{\mathrm{e\kern -0.1em V}}}
\newcommand{\GeVcnosp}{\ensuremath{{\mathrm{Ge\kern -0.1em V\!/}c}}}
\newcommand{\MeVcnosp}{\ensuremath{{\mathrm{Me\kern -0.1em V\!/}c}}}
\newcommand{\GeVccnosp}{\ensuremath{{\mathrm{Ge\kern -0.1em V\!/}c^2}}}
\newcommand{\MeVccnosp}{\ensuremath{{\mathrm{Me\kern -0.1em V\!/}c^2}}}
\newcommand{\GeVGeVccccnosp}{\ensuremath{{\mathrm{Ge\kern -0.1em V^2\!/}c^4}}}

\def\genxicctwo   {\mbox{\textsc{GenXicc2.0}}\xspace}

\def\XiccppDecay  {{\ensuremath{\Xiccpp\to\Lcp\Km\pip\pip}}\xspace}
\def\XiccppXicDecay  {{\ensuremath{\Xiccpp\to\Xicp\pip}}\xspace}

\newcommand{\nogapmevcc}{\ensuremath{{\mathrm{Me\kern -0.1em V\!/}c^2}}\xspace}

\def\RBF {{\ensuremath{\mathcal R(\BF)}}\xspace}
\def\XicpDecay  {{\ensuremath{\Xicp\to\proton\Km\pip}}\xspace}

\newcommand{\significance}{\nobreak{\ensuremath{ 5.9\sigma}}}
\newcommand{\rawm}{\nobreak{\ensuremath{ 3620.7\pm 1.5\mevcc}}}
\newcommand{\mass}{\nobreak{\ensuremath{ 3620.6\pm 1.5\stat\pm 0.4\syst \pm 0.3\, (\Xicp)\mevcc}}}
\newcommand{\ratio}{\nobreak{\ensuremath{ 0.035\pm 0.009\stat\pm 0.003\syst}}}
\newcommand{\syield}{\nobreak{\ensuremath{ 91\pm 20}}}
\usepackage{multirow}

\begin{document}

\renewcommand{\thefootnote}{\fnsymbol{footnote}}
\setcounter{footnote}{1}


\begin{titlepage}
\pagenumbering{roman}

\vspace*{-1.5cm}
\centerline{\large EUROPEAN ORGANIZATION FOR NUCLEAR RESEARCH (CERN)}
\vspace*{1.5cm}
\noindent
\begin{tabular*}{\linewidth}{lc@{\extracolsep{\fill}}r@{\extracolsep{0pt}}}
\ifthenelse{\boolean{pdflatex}}
{\vspace*{-1.5cm}\mbox{\!\!\!\includegraphics[width=.14\textwidth]{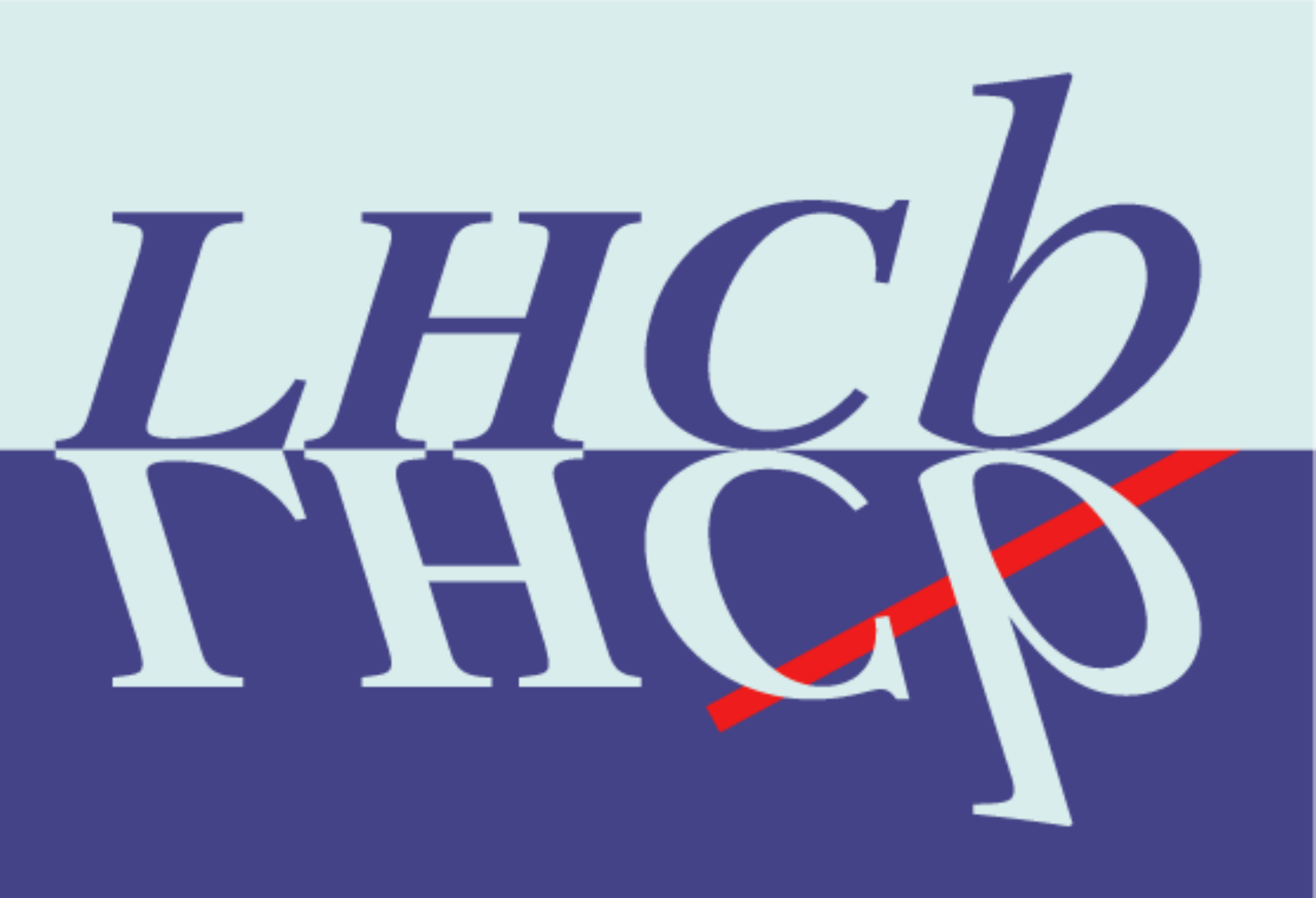}} & &}%
{\vspace*{-1.2cm}\mbox{\!\!\!\includegraphics[width=.12\textwidth]{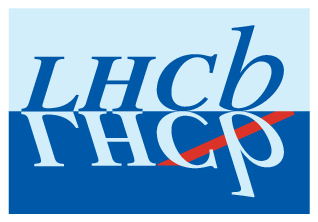}} & &}%
\\
 & & CERN-EP-2018-172 \\  
 & & LHCb-PAPER-2018-026 \\  
 & & October 18, 2018 \\ 
\end{tabular*}

\vspace*{4.0cm}

{\normalfont\bfseries\boldmath\huge
\begin{center}
  \papertitle 
\end{center}
}

\vspace*{2.0cm}

\begin{center}
\paperauthors\footnote{Authors are listed at the end of this paper.}
\end{center}

\vspace{\fill}

\begin{abstract}
  \noindent
  The doubly charmed baryon decay $\Xiccpp \to \Xicp \pip$
  is observed for the first time, with a statistical significance of
  $\significance$, confirming a recent observation of
  the baryon in the $\Lc \Km \pip \pip$ final state.
  The data sample used corresponds to an integrated luminosity of
  $1.7\invfb$, 
  collected by the \lhcb experiment in $pp$ collisions at a center-of-mass energy of $13\tev$. 
  The \Xiccpp mass is measured to be 
  \begin{equation}\nonumber
  \mass, 
  \end{equation}
  and is consistent with the previous result. 
  The ratio of branching fractions between the decay modes is measured to be
   \begin{equation}\nonumber
  \frac{
	\BF(\Xiccpp \to \Xicp \pip) \times \BF(\Xicp \to \proton \Km \pip)
  }{
	\BF(\Xiccpp \to \Lc \Km \pip \pip) \times \BF(\Lc \to \proton \Km \pip)
  } = \ratio.
  \end{equation}
\end{abstract}
\vspace*{2.0cm}

\begin{center}
  Published in Phys.~Rev.~Lett. {\bf 121}, 162002 (2018)
\end{center}

\vspace{\fill}

{\footnotesize 
\centerline{\copyright~\papercopyright. \href{\paperlicenceurl}{\paperlicence}.}}
\vspace*{2mm}

\end{titlepage}

\newpage
\setcounter{page}{2}
\mbox{~}

\cleardoublepage


\renewcommand{\thefootnote}{\arabic{footnote}}
\setcounter{footnote}{0}



\pagestyle{plain} 
\setcounter{page}{1}
\pagenumbering{arabic}


%

The recent observation by the \lhcb collaboration~\cite{LHCb-PAPER-2017-018} 
of a new state that is consistent with 
the doubly charmed baryon \Xiccpp,
opens a new field of research studying the properties 
of baryons containing two heavy quarks, providing a unique environment for
testing models of quantum chromodynamics (QCD).  
In studies of a sample of \Xiccpp decays to the final state $\Lc \Km \pip \pip$,
with $\Lc \to \proton \Km \pip$,
its mass was found to be 
\mbox{$3621.40\,\pm\,0.72 \stat \pm\,0.27\,\syst
\pm\,0.14\,{\,(\Lcp)}$\mevcc}~\cite{LHCb-PAPER-2017-018},
and its lifetime was measured to be 
$0.256\,^{+0.024}_{-0.022}\stat\pm\,0.014\,\syst$\ps~\cite{LHCb-PAPER-2018-019}.\footnote{
   The inclusion of charge-conjugate processes is implied throughout.
}
The measured lifetime firmly establishes its weakly decaying nature. 
Searching for new decay modes is the next critical step towards understanding
the dynamics of weak decays of doubly heavy baryons,
which may differ significantly from those of singly heavy hadrons
due to interference between decay amplitudes of the two heavy quarks. 
The process \XiccppXicDecay has been predicted to have a sizable 
branching fraction~\cite{Sharma:2017txj,Yu:2017zst},
making it a promising final state in which to seek confirmation of
	the previous observation.

\begin{figure}[bh]
\centering \includegraphics[scale=1.0]{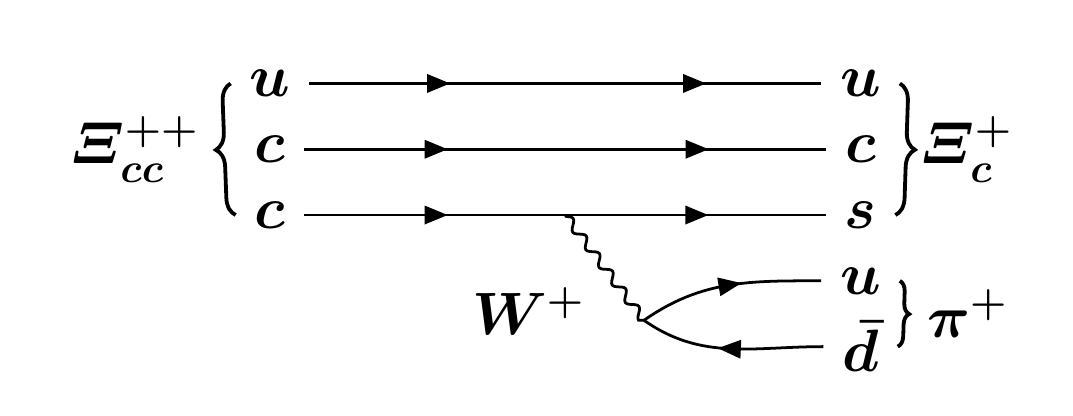}
\caption{Dominant Feynman diagram contributing to the decay \XiccppXicDecay.}
\label{fig:FeynmanDiagram}
\end{figure}

This Letter reports the first observation of 
the decay $\XiccppXicDecay$, 
which proceeds predominantly via the tree-level amplitude represented by the
Feynman diagram 
shown in Fig.~\ref{fig:FeynmanDiagram}.
The \Xicp baryon is reconstructed in its Cabibbo-suppressed decay 
to $\proton\Km\pip$.
The data sample used consists of $\proton\proton$ collisions
at a center-of-mass energy of 13\tev
collected by the \lhcb experiment in 2016,
corresponding to an integrated luminosity of $1.7$\invfb.
A measurement of the \Xiccpp mass with this sample is presented,
and the ratio of the total branching fractions, $\RBF$, between the decays
$\XiccppXicFullDecay$ and $\XiccppFullDecay$ is determined.

The \lhcb detector~\cite{Alves:2008zz,LHCb-DP-2014-002} is a
single-arm forward spectrometer covering the pseudorapidity range $2 < \eta < 5$, designed for
the study of particles containing \bquark\ or \cquark\ quarks. 
The detector includes a high-precision tracking system
consisting of
a silicon-strip vertex
detector~\cite{LHCb-DP-2014-001} surrounding the $pp$ interaction
region that allows \cquark\ and \bquark\ hadrons to be identified from
their typical long
flight distance; a tracking system~\cite{LHCb-DP-2013-003} that 
provides a measurement of momentum, $p$, of charged
particles; two ring-imaging Cherenkov detectors~\cite{LHCb-DP-2012-003} that discriminate between
different species of charged hadrons;
a calorimeter system consisting of
scintillating-pad and preshower detectors, an electromagnetic
calorimeter and a hadronic calorimeter, to identify photons, electrons and hadrons; 
a muon system composed of alternating layers of iron and multiwire
proportional chambers~\cite{LHCb-DP-2012-002} to identify muons.
The online event selection is performed by a trigger~\cite{LHCb-DP-2012-004}, 
which consists of a hardware stage, based on information from the calorimeter and muon
systems~\cite{LHCb-DP-2013-001}, 
followed by a software stage, which applies a full event
reconstruction incorporating real-time
alignment and calibration of the detector~\cite{LHCb-PROC-2015-011}.
The same alignment and calibration information is propagated to the offline reconstruction, ensuring consistent and high-quality particle identification (PID) information between the trigger and offline software.
The identical performance of the online and offline reconstruction offers the opportunity to perform physics analyses directly using candidates reconstructed in the trigger which is done in the present analysis.

Simulation is required to model the effects of 
the detector acceptance and the imposed selection requirements. 
In the simulation, $pp$ collisions are generated using
\pythia~\cite{Sjostrand:2007gs,*Sjostrand:2006za} 
with a specific \lhcb
configuration~\cite{LHCb-PROC-2010-056}.  
A dedicated generator, \genxicctwo~\cite{Chang:2007pp,*Chang:2009va}, 
is used to simulate \Xiccpp baryon production.
Decays of hadrons
are described by \evtgen~\cite{Lange:2001uf}, in which final-state
radiation is generated using \photos~\cite{Golonka:2005pn}. 
The interaction of the generated particles with the detector, and its response,
are modelled using the \geant
toolkit~\cite{Allison:2006ve, *Agostinelli:2002hh} as described in
Ref.~\cite{LHCb-PROC-2011-006}.

The selection of $\XiccppXicFullDecay$ decays is designed
to be as similar as possible to those of $\XiccppFullDecay$, 
described in Ref.~\cite{LHCb-PAPER-2017-018}. 
Three charged particles identified as $\proton$, $\Km$ and $\pip$ that form 
a good-quality vertex are combined to reconstruct a $\XicpDecay$ candidate.
The three particles are required to have transverse momenta in excess of $500\mevc$ and 
be inconsistent with originating from any primary vertex (PV).
The \Xicp vertex is required to be displaced from any PV by a distance corresponding to a \Xicp decay time greater than 0.15\ps,
which corresponds to approximately twice the decay time resolution.
The invariant-mass of each \Xicp candidate is required to be in the range 2450--2488\mevcc, corresponding to approximately six times the \Xicp mass resolution.
An additional positively charged particle, which must be identified as a
pion and have $\pt$ greater than 200\mevc, is then combined with
the $\Xicp$ candidate to form a \Xiccpp candidate.
The $\Xicp \pip$ pair is required to form a vertex that is of good quality and is upstream of the \Xicp vertex.
The $\Xiccpp$ candidate must have $\pt > 2000\mevc$ and be consistent with originating from a PV.
The candidate is associated with the PV with respect to which it has the smallest impact parameter $\chisq$ ($\chisqip$).
The $\chisqip$ is defined as the difference in $\chisq$ of the PV fit with and without the particle in question.
To avoid contributions due to duplicate tracks, candidates are rejected
if the angle between any pair of their final-state particles with the same charge
is smaller than $0.5$\mrad.
Specific hardware trigger requirements are also applied, 
to increase the signal yield and simplify the study of the trigger efficiency. 
Candidates are retained only if the event contains large transverse
energy deposits in the calorimeter arising from the decay products of the
\Xiccpp candidate, 
or if the event contains activity either in the calorimeter or in the muon system from particles other than these decay products.
Simulation shows that the efficiency for these additional requirements is above 90\% for both two-body or four-body \Xiccpp decay modes.

A multivariate selector based on the multilayer perceptron
algorithm~\cite{Hocker:2007ht,*TMVA4} is used to
further suppress combinatorial backgrounds.
To train the selector, simulated $\XiccppXicFullDecay$ decays 
are used as a signal sample,
and 0.3 million candidates from the upper sideband with invariant-masses in the range 
3800--4000\mevcc are used as a background sample.
To reduce the effect of the \Xicp mass resolution on the invariant-mass of the \Xiccpp candidates, an alternative
evaluation of the invariant-mass is used, ${m(\Xicp\pip)\equiv M(\Xicp\pip) - M([\proton\Km\pip]_{\Xicp}) + M_{\mathrm{PDG}}(\Xicp)}$, where
$M(\Xicp\pip)$ and $M([\proton\Km\pip]_{\Xicp})$ are the reconstructed masses of the \Xiccpp and \Xicp candidates, and
$M_{\mathrm{PDG}}(\Xicp)$ is the known value of the \Xicp mass~\cite{PDG2018}.

The input variables used in the multivariate selector are chosen based
on their discrimination power between signal and background
candidates. 
Three different types of variables are considered in the training. 
The first type of variables are the kinematic
information of particles, including the \pt of each of the four final-state
particles and of the \Xicp and \Xiccpp candidates; 
the angle between the \Xiccpp momentum vector and the displacement vector from the PV to the \Xiccpp decay vertex. 
The second type of variables are the vertex fitting qualities, including
the $\chisq$ per degree of freedom of the \Xicp and \Xiccpp vertex fits; 
the $\chisq$ per degree of freedom of a kinematic refit~\cite{Hulsbergen:2005pu} of the \XiccppXicFullDecay decay chain that requires the \Xiccpp to originate from its PV. 
The third type of variables are related to the lifetime, including
the $\chisqip$ of each of the four final-state particles and of the \Xicp
and \Xiccpp candidates with respect to their associated PV; 
the sum of the $\chisqip$ of the four final-state particles; 
and the flight distance $\chisq$ of the \Xicp and \Xiccpp candidates. 
The flight distance $\chisq$ is defined as the $\chisq$ of the hypothesis that 
the decay vertex of the candidate coincides with its associated PV.

Candidates are retained only if the multivariate-selector output exceeds a certain threshold.
This threshold is chosen to maximize the expected value of
the figure of merit
$\varepsilon/(\frac{5}{2}+\sqrt{N_{\rm B}})$~\cite{Punzi:2003bu}.
Here, $\varepsilon$ is the estimated signal efficiency and
$N_{\rm B}$ is the expected number of background candidates 
under the signal peak in the $\Xiccpp$ mass distribution,
after the selection.
The quantity $N_{\rm B}$ is determined, assuming an exponential shape for the background, from the number of $\Xicp\pip$
candidates in the mass region of 3800--4000\mevcc, 
scaled to a signal region 
centered at a mass of 3620\mevcc and with a width of 30\mevcc.
This corresponds to approximately five times the expected \Xiccpp mass resolution.
To test for potential biases in the multivariate selection
or other misreconstruction effects,
the same selection criteria are applied to control samples of data
consisting of $\Xicp\pip$ candidates in the \Xicp sideband regions
and of wrong-sign combination $\Xicp\pim$.
No peaking structure is visible in either samples.

\begin{figure}
\begin{center}
 \includegraphics[scale=0.42]{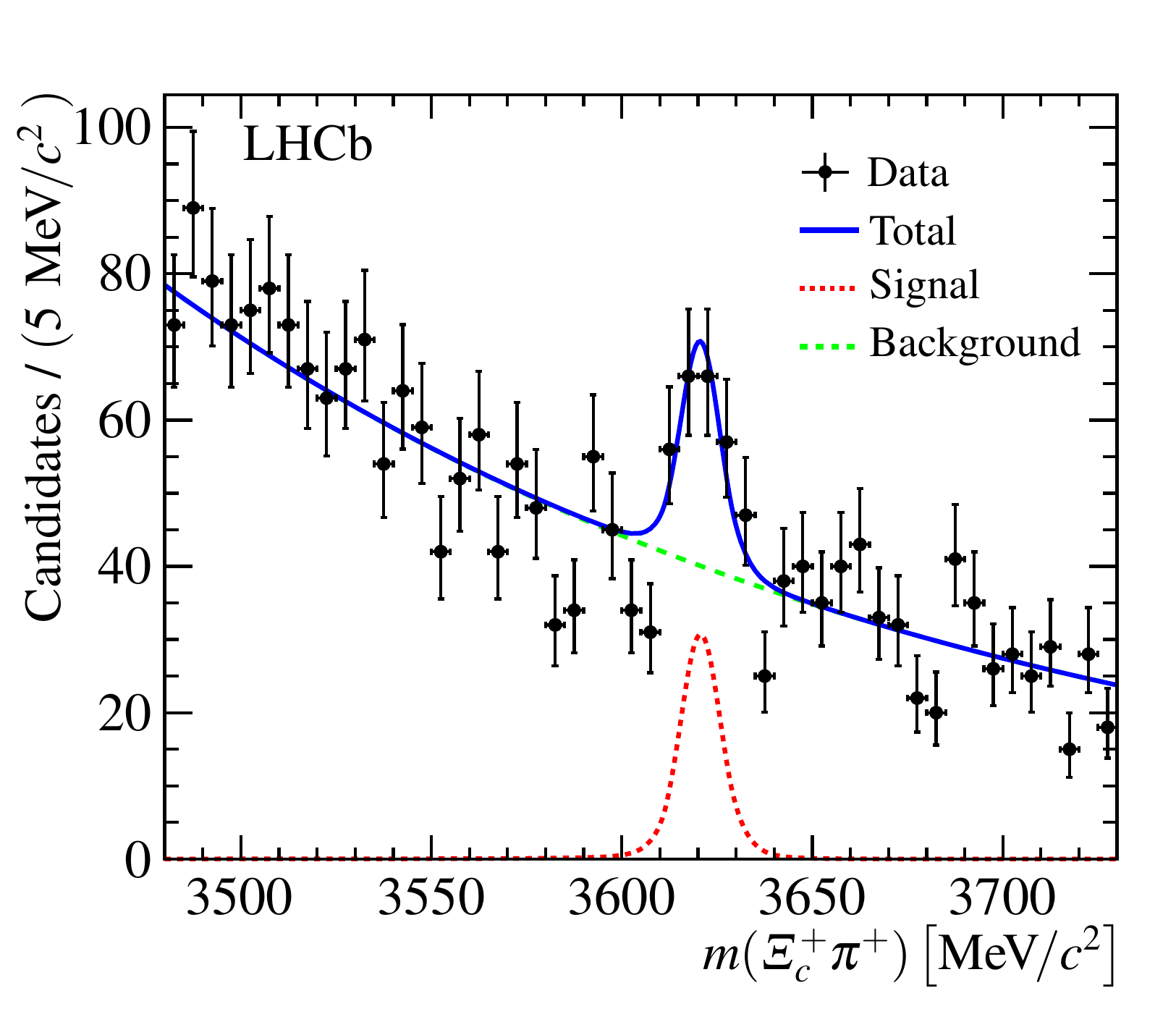}
 \includegraphics[scale=0.42]{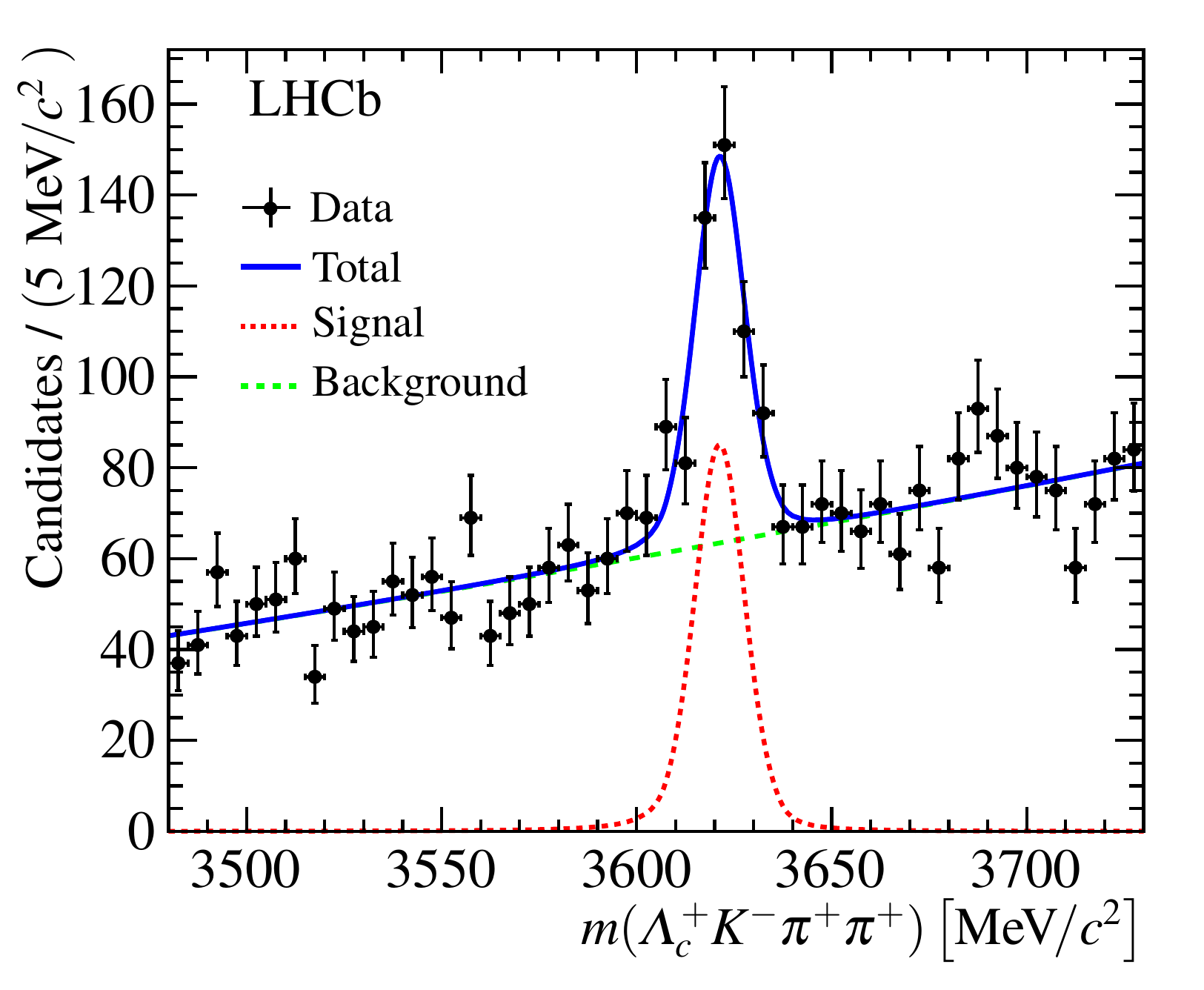}
\caption{Invariant-mass distribution of the $\XiccppXicDecay$ (left) and $\XiccppDecay$ (right)
  candidates 
with result of the fit overlaid. 
The black points represent the data, the dotted (red) line represents the
signal contribution, and the dashed (green) line represents the combinational background.}
\label{fig:massFit}
\end{center}
\end{figure}

Figure~\ref{fig:massFit} (left) shows the distribution of invariant-masses of
\Xiccpp candidates, $m(\Xicp\pip)$, after applying the complete selection. 
The contribution from events containing multiple signal candidates is found to be less than
1\%; all these candidates are included in the fit.
A signal is visible at a mass of approximately 3620\mevcc,
in the vicinity of the previous \lhcb~\Xiccpp baryon observation~\cite{LHCb-PAPER-2017-018}.
The mass distribution is fitted with an unbinned extended 
maximum-likelihood method to measure the properties of this structure. 
The peak is 
described by an empirical model, consisting of
a Gaussian function and
a modified Gaussian function with power-law tails on both sides~\cite{Skwarnicki:1986xj} and with the same mean value.
All tail parameters are fixed to values obtained from a fit to simulated signal events, 
while the parameters corresponding to the mass and the mass resolution are varied in the fit. 
The background shape is described by an exponential function.
The resulting signal yield is \syield\ and the mass value is $\rawm$,
where the uncertainties are statistical only.
The mass is fully consistent with the value measured 
in the $\XiccppDecay$ decay channel~\cite{LHCb-PAPER-2017-018},
and the resolution determined by the fit is consistent with expectations based on known detector performance.
The local statistical significance of the signal,
evaluated by taking the likelihood ratio 
corresponding to fits that include and exclude the signal component, 
is found to be $\significance$, thus confirming the observation reported in Ref~\cite{LHCb-PAPER-2017-018}.

The invariant-mass distribution for the reference mode, 
${\XiccppDecay}$, is shown in Fig.~\ref{fig:massFit} (right).
The selection used to obtain this sample is identical to that of the previous
analysis~\cite{LHCb-PAPER-2017-018}, except for the additional requirements on the 
hardware trigger.
An extended unbinned maximum-likelihood fit to the invariant-mass distribution
returns a signal yield of $289\pm35$ for the reference mode.

The branching fraction ratio, $\RBF$, between the decays $\XiccppXicFullDecay$ and $\XiccppFullDecay$ is defined as
\begin{eqnarray}
\label{eq:br}
  \RBF \equiv
  \frac{
    \BF \left( \Xiccpp \to \Xicp \pip \right)
    \times
    \BF \left( \Xicp \to \proton \Km \pip \right)
  }{
	\BF \left( \Xiccpp \to \Lc \Km \pip \pip \right)
    \times
    \BF \left( \Lc \to \proton \Km \pip \right)
  }
  =\frac{r_N}{r_{\eps}},
\end{eqnarray}
where $r_N$ is the ratio of \Xiccpp yields between the signal
and reference decay modes, and $r_{\eps}$ is the ratio of
total efficiencies between the two modes.
In each case, the total efficiency
includes the effects of the geometrical acceptance, trigger, 
reconstruction, and selection.
Each contribution to the efficiency ratio is evaluated with simulation, 
calibrated with data when possible, as described in the following.
The combined efficiency of the reconstruction and the selection,
excluding the hardware-trigger 
requirement, is determined from fully simulated signal samples in which 
the tracking~\cite{LHCb-DP-2013-002} and particle-identification efficiencies are 
corrected using control samples.
The correction to the efficiency ratio of the \XiccppDecay and \XiccppXicDecay
channels is determined to be $0.983\pm0.007$ for the tracking efficiency,
and $1.050\pm0.020$ for the particle-identification efficiency.
The hardware-trigger efficiency ratio is estimated from
fully simulated signal events, with a \pt-dependent correction derived from
data using ${\LbLcthreepi}$ and ${\LbLconepi}$ decays, 
which are required to pass the same trigger selection as the \Xiccpp candidates.
These two decay channels have similar final states and decay topologies as the signal.
The total relative efficiency
is determined to be $r_{\eps} = 0.110\pm0.002$, 
where the uncertainty comes from the limited size of the simulation sample and is accounted as a systematic uncertainty.
To validate this procedure, the ratio of branching fractions of the decays
${\LbLconepi}$ and ${\LbLcthreepi}$ is measured using the same
data sample, resulting in a value 0.83$\pm$0.05 (statistical uncertainty only) which agrees with 
the previous \lhcb result of 0.70$\pm$0.10~\cite{LHCb-PAPER-2011-016}.

The main sources of systematic uncertainty that affect the measurements
of the $\Xiccpp$ mass 
are summarized in Table~\ref{tab:MassMeasurementSys}.
Samples of $\jpsi\to\mumu$ and
$\Bp\to\jpsi\Kp$
decays~\cite{LHCb-PAPER-2011-035,LHCb-PAPER-2013-011}
are used to calibrate the reconstructed momentum of charged
particles, which affect the reconstructed mass of signal.
The maximum difference between the correction factors determined 
with above-mentioned decays
is found to be $0.03\%$,
which corresponds to a systematic uncertainty of $0.38\mevcc$ on the
measured \Xiccpp mass.
The signal selection efficiency increases with the \Xiccpp decay time; 
combined with a correlation between the reconstructed mass and 
the reconstructed decay time, this induces a positive bias on the masses of 
both $\Xiccpp$ and $\Xicp$ candidates.
The effect is studied with simulation and the bias 
on the measured \Xiccpp mass is found to be $+0.10 \pm 0.10$\mevcc, 
where the uncertainty is due to the limited size of the simulated samples.
A correction to the \Xiccpp mass of $-0.10$\mevcc is therefore applied,
and a systematic uncertainty of $0.10$\mevcc assigned.
The dependence of this bias on the \Xiccpp lifetime 
is studied by weighting simulated events to different lifetime hypotheses;
the change is found to be negligible for the measured 
\Xiccpp lifetime~\cite{LHCb-PAPER-2018-019}.
The description of the final-state radiation in simulation~\cite{Golonka:2005pn} 
can also cause a bias in the measured mass,
which is estimated with pseudoexperiments.
The correction is determined to be $+0.03$\mevcc,
with a negligible uncertainty.
The impact of the model used to fit the invariant-mass distribution 
on the measured mass is estimated by
varying the shape parameters that are fixed according to simulation,
using alternative signal and background models,
and performing the fits over different mass ranges.
The largest variation in the fitted $\Xiccpp$ masses, $0.05\mevcc$,
 is taken as a systematic uncertainty.
The current known value for the \Xicp mass~\cite{PDG2018} 
is used to compute the invariant-mass $m(\Xicp\pip)$ of the $\Xiccpp$
candidate. Its uncertainty, $0.30$\mevcc,
is assigned as a systematic uncertainty on the $\Xiccpp$ mass. 

The systematic uncertainties on the ratio $\RBF$ are listed in
Table~\ref{tab:MassMeasurementSys} and are described as follows.
The alternative fit models mentioned above result in different values of the ratio $r_N$. The largest relative deviation measured, $5.2\%$, is assigned as a systematic uncertainty on $\RBF$.
The relative efficiency of the tracking, particle identification and 
trigger are estimated using control samples, 
whose statistical uncertainties are taken as
a systematic uncertainty on $\RBF$.
An additional uncertainty of 4.1\% is assigned 
on the track-reconstruction efficiency due to uncertainties on 
the material budget of the detector and the modelling of hadronic interaction
with the detector material. 
The {particle-identification} efficiency is determined in bins of particle momentum and pseudorapidity using control samples.
The size of the bins is increased or decreased by a factor of two and the largest deviation on $\RBF$ is assigned as systematic uncertainty related to the binning.
An additional uncertainty of 4.2\% on the hardware trigger efficiency is determined from
the \Lb control samples described above, 
including a statistical uncertainty from the limited sample size,
and an uncertainty that is determined by testing the procedure in simulation and taking the deviation as a systematic uncertainty.
Combining the systematic uncertainties on the efficiency mentioned above, a systematic uncertainty of $6.5\%$ on \RBF is assigned.
Uncertainties from the \Xiccpp mass, 
lifetime and production spectra are investigated, 
and 1.2\% is assigned as a systematic uncertainty.
Different requirements on the \Xiccpp \pt are applied 
to select the $\XiccppXicDecay$ and $\XiccppDecay$ decays, 
and this may cause a bias if the \pt
distribution of simulated \Xiccpp differs from that in data.
To assess the size of this effect,
the measurement is repeated 
applying the same \pt requirement to both modes.
The difference in \RBF is found to be $0.7\%$.
A separate measurement carried out with a cut-based selection gives a consistent result.

\begin{table}
  \caption{
      Systematic uncertainties on the measurement of the $\Xiccpp$
      mass and of the ratio of branching fractions \RBF between the 
      $\XiccppXicDecay$ and the $\XiccppDecay$ decay modes.
  }
\begin{center}
\begin{tabular}{lcc}
\hline
Source                       & Mass $[\mevcc]$   & ${\cal{R}}(\BF)$ [\%]\\
\hline
 Momentum calibration        & 0.38          & --- \\
 Selection bias correction   & 0.10          & --- \\
 Fit model              & 0.05          & 5.2 \\
 Relative efficiency         & ---             & 6.5 \\
 Simulation modelling  & ---             & 1.2 \\
 Selection                      & ---                & 0.7 \\
\hline
Sum in quadrature            & 0.40          & 8.5 \\
\hline
\end{tabular}\label{tab:MassMeasurementSys}
\end{center}
\end{table}

The value of $\RBF$ is measured to be $\ratio$ and
the \Xiccpp mass is measured to be \mass, which is 
consistent with the mass measured in the final state
$\Lc \Km \pip \pip$,
\mbox{$3621.40\,\pm\,0.72 \stat \pm\,0.27\,\syst
\pm\,0.14\,{\,(\Lcp)}$\mevcc}~\cite{LHCb-PAPER-2017-018}.
Averaging over the two measurements,
the \Xiccpp mass is determined to be $3621.24\pm0.65\stat\pm0.31\syst\mevcc$
(see the Appendix~\ref{sec:Supplementary-App} for the comparisons between the measured \Xiccpp masses and combined result).
The combination is performed using the Best Linear Unbiased Estimate
(BLUE) method~\cite{Lyons:1988rp,Valassi:2003mu}.
In the combination, the systematic uncertainties are assumed to be
uncorrelated except for the momentum scale calibration.

In summary, a new decay mode of the doubly charmed baryon
$\XiccppXicDecay$ is observed with a statistical significance of $\significance$
in a data sample of $\proton\proton$ collisions
collected by the \lhcb experiment
at a center-of-mass energy of $\sqs=13$\tev.
The \Xiccpp mass is consistent with the previous \lhcb result~\cite{LHCb-PAPER-2017-018} and with most theoretical calculations of the \Xiccpp mass
(see \eg Ref.~\cite{Alexandrou:2017xwd}).
The ratio of the total branching fractions between this decay (\XiccppXicDecay) and
the reference mode (\XiccppDecay) is 
consistent with the prediction of Ref.~\cite{Yu:2017zst}, 
which, however, has large uncertainties.
Therefore, this measurement provides important information
towards an improved understanding of the decays of
doubly charmed baryons.

\section*{Acknowledgements}
%
%
\noindent We thank Chao-Hsi Chang, Cai-Dian L\"u, Wei Wang, Xing-Gang Wu, and
Fu-Sheng Yu for frequent and interesting discussions on the production
and decays of double-heavy-flavor baryons.
We express our gratitude to our colleagues in the CERN
accelerator departments for the excellent performance of the LHC. We
thank the technical and administrative staff at the LHCb
institutes.
We acknowledge support from CERN and from the national agencies:
CAPES, CNPq, FAPERJ and FINEP (Brazil); 
MOST and NSFC (China); 
CNRS/IN2P3 (France); 
BMBF, DFG and MPG (Germany); 
INFN (Italy); 
NWO (Netherlands); 
MNiSW and NCN (Poland); 
MEN/IFA (Romania); 
MinES and FASO (Russia); 
MinECo (Spain); 
SNSF and SER (Switzerland); 
NASU (Ukraine); 
STFC (United Kingdom); 
NSF (USA).
We acknowledge the computing resources that are provided by CERN, IN2P3
(France), KIT and DESY (Germany), INFN (Italy), SURF (Netherlands),
PIC (Spain), GridPP (United Kingdom), RRCKI and Yandex
LLC (Russia), CSCS (Switzerland), IFIN-HH (Romania), CBPF (Brazil),
PL-GRID (Poland) and OSC (USA).
We are indebted to the communities behind the multiple open-source
software packages on which we depend.
Individual groups or members have received support from
AvH Foundation (Germany);
EPLANET, Marie Sk\l{}odowska-Curie Actions and ERC (European Union);
ANR, Labex P2IO and OCEVU, and R\'{e}gion Auvergne-Rh\^{o}ne-Alpes (France);
Key Research Program of Frontier Sciences of CAS, CAS PIFI, and the Thousand Talents Program (China);
RFBR, RSF and Yandex LLC (Russia);
GVA, XuntaGal and GENCAT (Spain);
Herchel Smith Fund, the Royal Society, the English-Speaking Union and the Leverhulme Trust (United Kingdom);
Laboratory Directed Research and Development program of LANL (USA).

\addcontentsline{toc}{section}{References}
\setboolean{inbibliography}{true}
\bibliographystyle{LHCb}
\bibliography{main,LHCb-PAPER,LHCb-CONF,LHCb-DP,LHCb-TDR,xicc-refs}

\newpage

\newpage
\centerline{\large\bf LHCb collaboration}
\begin{flushleft}
\small
R.~Aaij$^{27}$,
B.~Adeva$^{41}$,
M.~Adinolfi$^{48}$,
C.A.~Aidala$^{73}$,
Z.~Ajaltouni$^{5}$,
S.~Akar$^{59}$,
P.~Albicocco$^{18}$,
J.~Albrecht$^{10}$,
F.~Alessio$^{42}$,
M.~Alexander$^{53}$
% ...

{\footnotesize \it
% ...
$ ^{5}$Clermont Universit{\'e}, Universit{\'e} Blaise Pascal, CNRS/IN2P3, LPC, Clermont-Ferrand, France\\
% ...
$ ^{10}$Fakult{\"a}t Physik, Technische Universit{\"a}t Dortmund, Dortmund, Germany\\
% ...
$ ^{18}$INFN Laboratori Nazionali di Frascati, Frascati, Italy\\
% ...
$ ^{27}$Nikhef National Institute for Subatomic Physics, Amsterdam, Netherlands\\
% ...
$ ^{41}$Instituto Galego de F{\'\i}sica de Altas Enerx{\'\i}as (IGFAE), Universidade de Santiago de Compostela, Santiago de Compostela, Spain\\
$ ^{42}$European Organization for Nuclear Research (CERN), Geneva, Switzerland\\
% ...
$ ^{48}$H.H. Wills Physics Laboratory, University of Bristol, Bristol, United Kingdom\\
% ...
$ ^{53}$School of Physics and Astronomy, University of Glasgow, Glasgow, United Kingdom\\
% ...
$ ^{59}$University of Cincinnati, Cincinnati, OH, United States\\
% ...
$ ^{61}$Syracuse University, Syracuse, NY, United States\\
% ...
$ ^{73}$University of Michigan, Ann Arbor, United States, associated to $^{61}$\\
% ...
}
\end{flushleft}



\clearpage

\section{Supplementary material for LHCb-PAPER-2018-026}
\label{sec:Supplementary-App}

\begin{figure}[!h]
\begin{center}
 \includegraphics[scale=0.8]{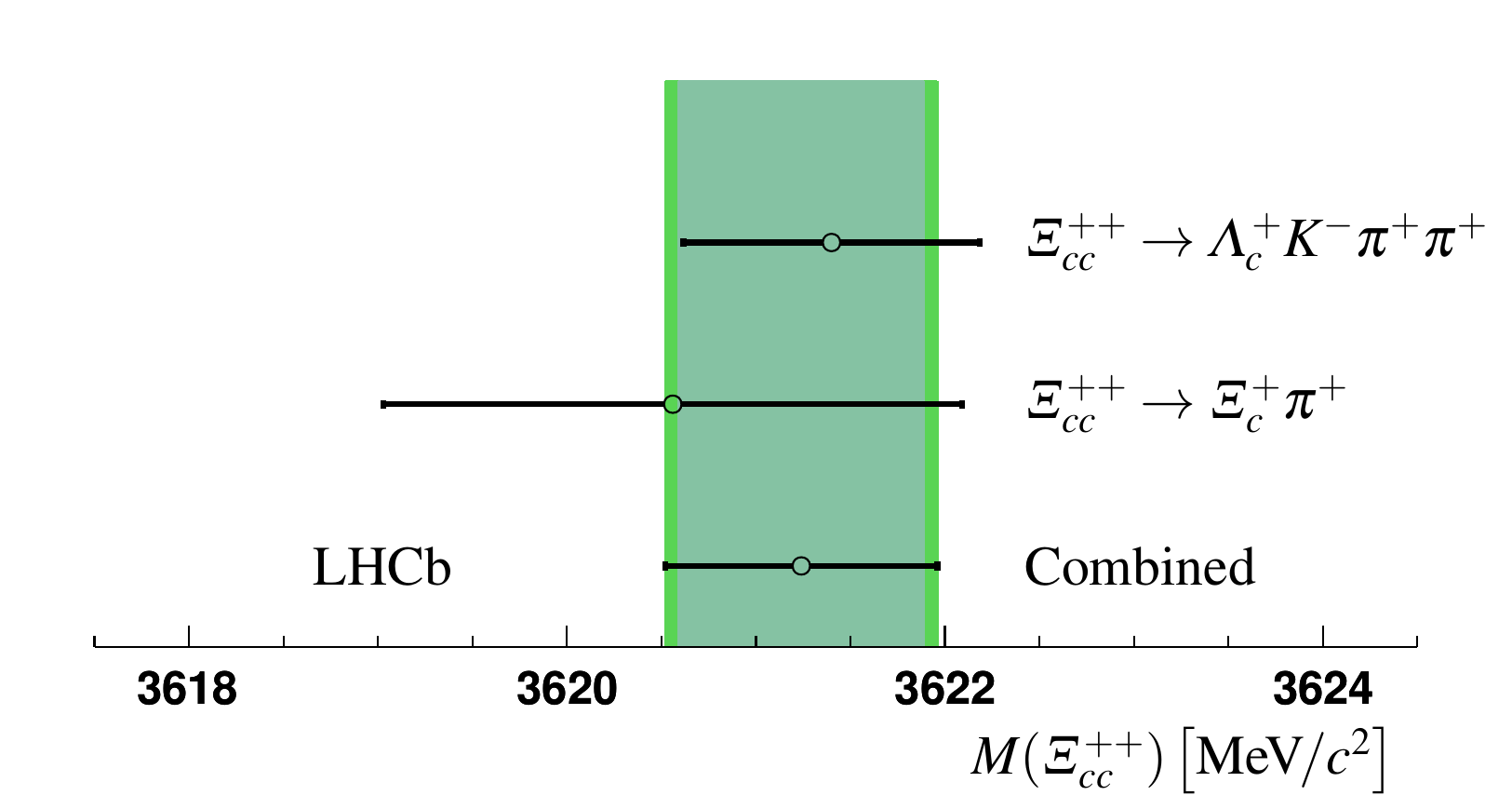}
\caption{Measured \Xiccpp masses obtained with the decay modes \XiccppDecay and \XiccppXicDecay,
and the combined result. The darker green band represents the statistical uncertainty on the combination, and the lighter green band represents the total uncertainty on the combination.}
\label{fig:masscom}
\end{center}
\end{figure}

Figure~\ref{fig:masscom} shows the comparison between \Xiccpp mass measured from 
\XiccppDecay and \XiccppXicDecay channel, and the combined
mass value using these two results.

\clearpage

\end{document}